\documentclass[conference]{IEEEtran}
\ifCLASSINFOpdf
   \usepackage[pdftex]{graphicx}
  \graphicspath{{CAMSAP_figures/}}
\else
   \usepackage[dvips]{graphicx}
   \graphicspath{CAMSAP_figures/}
\fi
\usepackage{multicol, blindtext}
\usepackage{amsmath}
\usepackage{enumerate}
\begin{document}
\title{Kronecker Sum Decompositions of Space-Time Data                      }

\author{\IEEEauthorblockN{Kristjan Greenewald}
\IEEEauthorblockA{University of Michigan \\ Ann Arbor}\and \IEEEauthorblockN{Theodoros Tsiligkaridis}
\IEEEauthorblockA{University of Michigan \\ Ann Arbor}\and \IEEEauthorblockN{Alfred O. Hero III}
\IEEEauthorblockA{University of Michigan \\ Ann Arbor}
}
  \maketitle
\begin{abstract}
In this paper we consider the use of the space vs. time Kronecker product decomposition in the estimation of covariance matrices for spatio-temporal data. This decomposition imposes lower dimensional structure on the estimated covariance matrix, thus reducing the number of samples required for estimation. To allow a smooth tradeoff between the reduction in the number of parameters (to reduce estimation variance) and the accuracy of the covariance approximation (affecting estimation bias), we introduce a diagonally loaded modification of the sum-of-kronecker products representation in \cite{TsiliArxiv}. 
We derive 
an asymptotic Cram\'{e}r-Rao bound (CRB) on the minimum attainable mean squared predictor coefficient estimation error for unbiased estimators of Kronecker structured covariance matrices. We illustrate the accuracy of the diagonally loaded Kronecker sum decomposition by applying it to the prediction of human activity video.
\end{abstract}

\section{Introduction}


In this paper, we develop a method for estimation of spatio-temporal covariance and apply it to video modeling and prediction. The covariance for spatio-temporal processes manifests itself as multiframe covariance, i.e. the covariance not only between pixels or features in a single frame, but also between pixels or features in a set of nearby frames. In streaming applications, at each time $t$ the covariance may be estimated over a sliding time window of $T$ frames. If each frame contains $N$ spatial components, e.g., pixels, then the covariance is described by a $NT$ by $NT$ matrix:
\begin{equation}
\mathbf{\Sigma}_t = \mathrm{Cov}\left[\{\mathbf{I_n}\}_{n=t-T}^{t-1}\right]
\end{equation}
where $\mathbf{I}_n$ denotes the $N$ pixels or other features of interest in the $n$th video frame. We make the standard piecewise stationarity assumption that $\mathbf{\Sigma}_t$ can be approximated as unchanging over each consecutive set of $T$ frames.

%

As $NT$ can be very large, even for moderately large $N$ and $T$ the number of degrees of freedom ($NT(NT+1)/2$) in the covariance matrix can greatly exceed the number $n$ of i.i.d. samples available to estimate the covariance matrix. One way to handle this problem is to introduce structure and/or sparsity into the covariance matrix, thus reducing the number of parameters to be estimated. In many spatio-temporal applications it is expected (and confirmed by experiment) that significant sparsity exists in the inverse pixel correlation matrix due to Markovian relations between neighboring pixels and frames. Sparsity alone, however, is not sufficient, and applying standard sparse methods such as GLasso directly to the spatio-temporal covariance matrix is computationally prohibitive \cite{tsiligkaridis2013convergence}.

A natural non-sparse alternative is to introduce structure is by modeling the covariance matrix $\mathbf{\Sigma}$ as the Kronecker product of two smaller matrices, i.e.
\begin{equation}
\label{KronApprox}
\mathbf{\Sigma} \approx \mathbf{T}\otimes \mathbf{S}.
\end{equation}
When  the measurements are Gaussian with covariance of this form they are said to follow a matrix-normal distribution \cite{tsiligkaridis2013convergence}. This model lends itself to coordinate decompositions \cite{tibshirani,TsiliArxiv,genton2007separable}. For spatio-temporal data, we consider the natural decomposition of space (features) vs. time (frames) \cite{TsiliArxiv,genton2007separable}. In this setting, the $\mathbf{S}$ matrix is the ``spatial covariance" and $\mathbf{T}$ is the ``time covariance."


Previous applications of the model of Equation \eqref{KronApprox} include MIMO wireless channel modeling as a transmit vs. receive decomposition \cite{werner2007estimation}, geostatistics \cite{Cressie1993}, genomics \cite{yin2012model}, multi-task learning \cite{bonilla2008multi}, collaborative filtering \cite{yu2009large}, face recognition \cite{zhang2010learning}, mine detection \cite{zhang2010learning}, and recommendation systems \cite{tibshirani}.

An extension to the representation \eqref{KronApprox} introduced in \cite{TsiliArxiv} approximates the covariance matrix using a sum of Kronecker product factors
\begin{equation}
\label{SumApprox}
\mathbf{\Sigma} \approx \sum\nolimits_{i=1}^{r} \mathbf{T}_i \otimes \mathbf{S}_i
\end{equation}
where $r$ is the separation rank.

This allows for more accurate approximation of the covariance when it is not in Kronecker product form but most of its energy is in the first few Kronecker components. An algorithm for fitting the model \eqref{SumApprox} to a measured sample covariance matrix was introduced in \cite{TsiliArxiv}. 
The Kronecker sum model does not naturally accommodate additive noise since the diagonal elements (variances) must conform to the Kronecker structure.

In this paper, we extend the Kronecker sum model, and the PRLS algorithm of \cite{TsiliArxiv}, by adding a structured diagonal matrix to \eqref{SumApprox}. This model is called the Diagonally Loaded Kronecker Sum model and, although it has an additional $N$ parameters, we show that it does significantly better at predicting video data. We also derive the asymptotic Cram\'{e}r-Rao lower bound on the estimation MSE of the ML predictor coefficients using both standard covariance and Kronecker estimation.




The rest of this paper is organized as follows. 
Section \ref{Sec:Pred} introduces the diagonally loaded Kronecker sum model and a LS algorithm for its estimation. We also derive the CRB based asymptotic predictor performance gain when estimating $\Sigma$ using the model of \eqref{KronApprox}. In section \ref{Sec:Est} we present results on the accuracy of the multiple Kronecker representation of real-world video spatio-temporal covariances. Section \ref{Sec:Results} presents our comparative prediction performance results using CMU activity video data, and we conclude the paper in Section \ref{Conclusion}.

\section{Sums-of-Kronecker Covariance representation for Prediction}
\label{Sec:Pred}
As an example application of covariance estimation, we turn in this section to the use of estimated spatio-temporal covariance matrices for prediction tasks. Given a covariance matrix $\mathbf{\Sigma}$ and mean $\mu$ of a vector with non-overlapping subvectors $x$ and $y$ the ML predictor of $y$ given $x$ is
\begin{equation}
\hat{y} = \mathbf{\Sigma}_{yx}\mathbf{\Sigma}_x^{-1} (x-\mu_x) + \mu_y
\label{Eq:Predictor}
\end{equation}
where $\mathbf{\Sigma}_{yx}$ and $\mathbf{\Sigma}_x$ are the appropriate submatrices of $\mathbf{\Sigma}$ \cite{bonilla2008multi}.


\subsection{Modified LS Algorithm for Prediction Tasks}





Although the Kronecker structure of video space-time covariance matrices is strong, the diagonal elements of any covariance matrix are strongly affected by any uncorrelated noise in the system \cite{bonilla2008multi}, which does not replicate across the matrix in a Kronecker fashion. Hence, for example, the Kronecker estimate will overestimate positive in-frame correlations.
Since the diagonal elements of a covariance matrix are highly important for determining the inverse of the matrix and by extension the predictor coefficients, this can cause a significant loss of accuracy. 

To correct this problem, we thus propose to approximate the covariance using the $r+1$-Kronecker model
\begin{equation}
\mathbf{{\Sigma}} \approx \left(\sum\nolimits_{i = 1}^{r}\mathbf{T}_i \otimes \mathbf{S}_i\right) + \mathbf{I} \otimes \mathbf{U}
\end{equation}
where $\mathbf{U}$ is diagonal \cite{bonilla2008multi}.


Since the diagonal addition is arbitrary, it does not matter what values the Kronecker portion assigns to the diagonal elements. Hence we set them as don't cares in the least-squares low separation rank approximation. We thus turn to the estimation of $\mathbf{T}$ and $\mathbf{S}$ from the sample covariance matrix $\mathbf{R}$ with the diagonal elements of $\mathbf{T}\otimes\mathbf{S}$ being don't cares. 

Following rearrangement of $\mathbf{R}$ to form $\mathbf{B}$ as in \cite{werner2008estimation}, this becomes the problem of finding a rank-one (low rank for multiple Kroneckers) approximation to a matrix $\mathbf{B}$ where the intersections of a set of rows and columns are not included in the LS objective function \cite{werner2008estimation}.

For notational simplicity, multiply $\mathbf{B}$ by permutation matrices to put it in the form
\begin{equation}
\mathbf{\tilde{B}} = \left[\begin{array}{cc} \mathbf{B}_{11} & \mathbf{B}_{12} \\ \mathbf{B}_{21} & \mathbf{B}_{22}\end{array}\right]
\end{equation}
where the don't cares are now contained in the ($T\times N$) $\mathbf{B}_{22}$. We also divide the permuted rank $r$ approximation matrices $\mathbf{t}$ and $\mathbf{s}$ in the same way, that is ${t}_i = \left[\begin{array}{cc} {t}_{i1}^T & {t}_{i2}^T \end{array}\right]^T \: \mathrm{and} \: {s}_i = \left[\begin{array}{cc} s_{i,1}^T & s_{i,2}^T \end{array}\right]^T $ where $t_i,s_i$ are the columns of $\mathbf{t},\mathbf{s}$.
As shown in \cite{werner2008estimation}, the vectors $\mathbf{t_i,s_i}$ can be rearranged to form the Kronecker factors $\mathbf{T_i,S_i}$ respectively. We thus have
\begin{equation}
\{\mathbf{\hat{t}},\hat{\mathbf{s}}\} = \arg \min_{\mathbf{t},\mathbf{s}} \|\mathbf{t} \mathbf{s}_1^T - \mathbf{B}_{1}\|_F^2 + \|\mathbf{t}_1 \mathbf{s}_2^T - \mathbf{B}_{12}\|_F^2,
\label{eq:OF}
\end{equation}
where $\mathbf{B}_{1} = [\mathbf{B}_{11};\mathbf{B}_{21}]$. Our algorithm is then:

1) Rearrange $\mathbf{R}$ to form $\mathbf{\tilde{B}}$.

2) Solve the (biconvex) weighted LS rank $r$ approximation problem in Equation \eqref{eq:OF}. We use the iterative method of alternating projections \cite{buchanan2005damped} over $\mathbf{t}$ and $\mathbf{s}$, initializing using the unweighted SVD solution since the number of missing values is relatively small ($NT$ out of $N^2T^2$). 

3) Reform $\mathbf{t}_i,\mathbf{s}_i$ to get $\hat{\mathbf{T}}_i, \hat{\mathbf{S}}_i$.

4) Determine $\mathbf{U}$, which must be diagonal. We set $u_{ii} = \mathbf{max}\left\{0,R_{ii}-\tilde{R}_{ii}\right\}$, where $\tilde{\mathbf{R}} = \sum_{i=1}^r\mathbf{T}_i\otimes \mathbf{S}_i$ and the zero cutoff exists as it helps preserve positive semidefiniteness. Further additions to $\mathbf{U}$ can be added for regularization.

We found that prediction accuracy is typically better if the diagonally loaded LS approximation is applied to the sample correlation instead of the sample covariance. 

\subsection{Cram\'{e}r-Rao Bound (CRB) on Predictor Coefficients}
The CRB on the asymptotic optimal performance of an unbiased estimator of a Kronecker product covariance matrix $\mathbf{\Sigma}= \mathbf{T} \otimes \mathbf{S}$ using $N$ iid samples is given by \cite{werner2008estimation}
\begin{align}
\label{eq:CRB}
N\mathrm{Cov}&[\mathrm{vec}\{\hat{\mathbf{\Sigma}}\}]\geq \mathbf{F}_{\Sigma} \\\nonumber
&= \mathbf{P}\mathbf{\Gamma}_0(\mathbf{\Gamma}_0^T\mathbf{P}^H(\mathbf{\Sigma}^{-T}\otimes \mathbf{\Sigma}^{-1}) \mathbf{P}\mathbf{\Gamma}_0)^\dag \mathbf{\Gamma}_0^T\mathbf{P}^H
\end{align}
where
\begin{align}
\mathbf{\Gamma}_0 = [\theta_S \otimes \mathbf{I}_{n_T\times n_T} \quad \mathbf{I}_{n_S \times n_S}\otimes \theta_T], \quad \mathbf{P} = \mathbf{P}_R (\mathbf{P}_S \otimes \mathbf{P}_T).\nonumber
\end{align}
$\mathbf{P}_R$ is a permutation matrix described in \cite{werner2008estimation},
and
$\theta_S,\theta_T,\mathbf{P}_S,\mathbf{P}_T$ are such that $\mathrm{vec}\{\mathbf{T}\} = \mathbf{P}_T\theta_T, \mathrm{vec}\{\mathbf{S}\} = \mathbf{P}_S\theta_S$ (allowing for imposition of certain types of structure).

The predictor coefficients are $\mathbf{A} = \mathbf{\Sigma}_{yx}\mathbf{\Sigma_x}^{-1}$.
Let $\mathbf{a} = \mathrm{vec}\{\mathbf{A}\}$. Then the asymptotic CRB of $\mathbf{a}$ is
\begin{align}
\label{Eq:PredCRB}
N\mathrm{Cov}&[\mathrm{vec}\{\hat{\mathbf{A}}\}]\geq \mathbf{F}_a \rightarrow \mathbf{J}^T \mathbf{F}_{\Sigma} \mathbf{J}, \: N \rightarrow \infty
\end{align}
where $\mathbf{J}$ is the Jacobian of $\mathbf{a}$ with respect to $\mathrm{vec}\{\mathbf{\Sigma}\}$. The values of $\mathbf{J}$ for the portions of $\mathbf{\Sigma}$ not used in the predictor coefficients are trivially zero. For the $\mathbf{\Sigma}_{yx}$ portion,
\begin{align}
\frac{\partial A_{ij}}{\partial [\mathbf{\Sigma}_{yx}]_{k\ell}} = [\mathbf{\Sigma}_x^{-1}]_{\ell j}\quad \forall k = i, \quad 0 \: o.w.
\end{align}
For the $\mathbf{\Sigma}_x$ portion,
\begin{align}
\frac{\partial A_{ij}}{\partial [\mathbf{\Sigma}_{x}]_{k\ell}} = -[\mathbf{\Sigma}_{x}^{-1}]_{\ell j} \sum\nolimits_f [\mathbf{\Sigma}_{yx}]_{if} [\mathbf{\Sigma}_{x}^{-1}]_{fk}.
\end{align}

Now that the CRB has been derived for the predictor coefficients, it is possible to obtain the asymptotic reduction in accuracy of the Kronecker based predictor $\hat{y}$ relative to the infinite training sample predictor. Assume that $x,y$ are independent of the training samples. Without loss of generality, assume $E[x]= 0$. Define
\begin{equation}
e = \hat{y} - E[y|x] = (\hat{\mathbf{A}} - \mathbf{A})x.
\end{equation}
Thus $E[e] = 0$. Also, by independence, $\mathrm{Cov}[\hat{y}-y] = \mathrm{Cov}[e] + \mathrm{Cov}[y|x]$. Since the CRB assumes an unbiased estimator, assume that the estimator of $\mathbf{A}$ is unbiased. Then $E[\hat{\mathbf{A}}]=E[\mathbf{A}]$. The error covariance is then given by
\begin{align}
\label{eq:ErrCov}
\mathrm{Cov}[e_i,e_j] &= E[(\hat{A}_i-A_i)x(\hat{A}_j -A_j)x] \\\nonumber&= \sum\nolimits_{k,\ell} \mathrm{Cov}[\hat{A}_{ik},\hat{A}_{j\ell}] \mathbf{\Sigma}_{x,k\ell}
\end{align}
where $A_i$ denotes the $i$th row of $\mathbf{A}$. The asymptotic covariance of the predictor coefficient estimates is given by the CRB for the predictor coefficients \eqref{Eq:PredCRB}, thus giving the asymptotic lower bound on the covariance of the additional error $e$ resulting from the use of the estimated instead of the true $\mathbf{\Sigma}$.

For comparison, the asymptotic CRB for covariance estimation (arbitrary $\mathbf{\Sigma}$) with no structural knowledge can be obtained by setting $\mathbf{T} = 1, \mathbf{S} = \mathbf{\Sigma}$ in Equation \eqref{eq:CRB}.

\section{Sum of Kronecker Products Decomposition Accuracy in Video}
\label{Sec:Est}
In this section, we focus on the MSE accuracy of approximating sample covariance matrices using the sum-of-Kroneckers approximation, in particular, the number of Kronecker sums required to obtain a good approximation. Since we are focusing on the MSE, the standard sum-of-kroneckers approximation of Equation \eqref{SumApprox} is appropriate.

\subsubsection{Texture Video}

For the computation of the sample covariance, we use a sliding window approach, where to obtain each new multiframe sample, the window is incremented by one frame. Linear dependence is avoided, but the samples are clearly not independent. However, it does improve the learning rate and enforces stationarity which could otherwise be lost in periodic situations.

\begin{figure}[htb]
\centering
\includegraphics[width=1.3in]{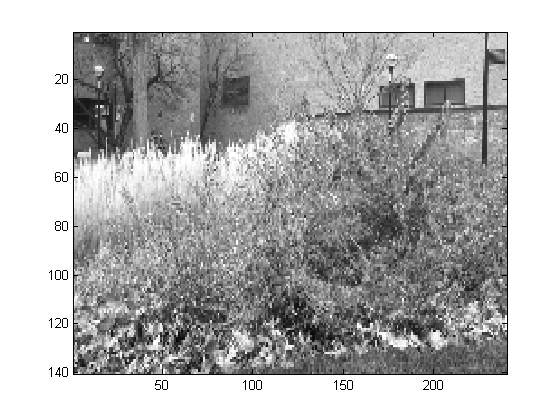}\includegraphics[width=1.3in]{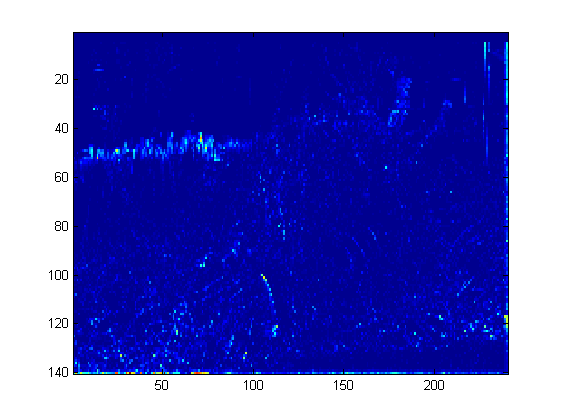}\includegraphics[width=1.3in]{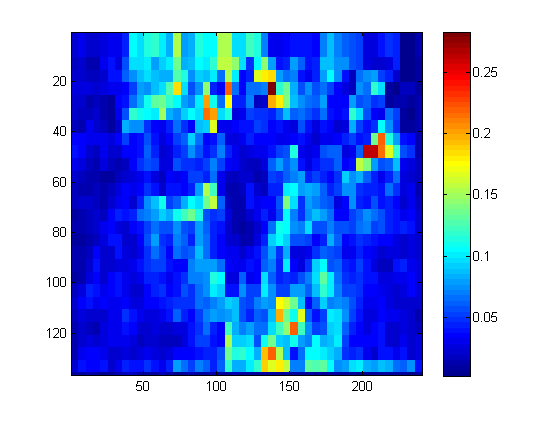}
\caption{Grass Video: still image, pixel variance image, and single Kronecker \% RMSE for 30 frame covariance at each block.}
\label{Fig:Grass}
\end{figure}

We examine the accuracy of the Kronecker approximation for a video of blowing grass (15 frames/sec). Due to the size of the image, it is divided up into blocks for estimation. Figure \ref{Fig:Grass} shows a still frame of the video, along with the variance image for reference. The \% RMSE of using a single Kronecker factor to approximate the 15 frame, 2000 sample, sample covariances is shown for each $10\times10$ block. Due to the relatively low number of samples compared to the number of variables (1500), a significant portion of this error is likely due to noise in the sample covariance.

\begin{figure}[htb]
\centering
\includegraphics[width=1.2in]{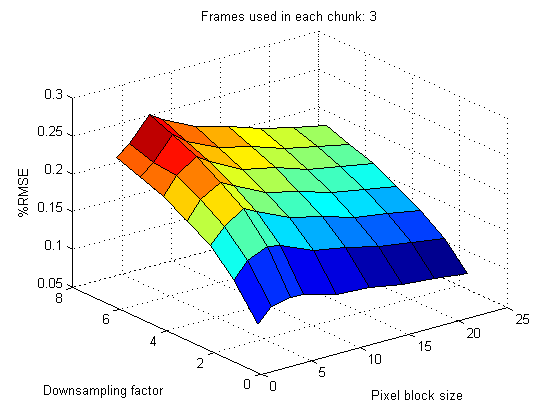}\includegraphics[width=1.2in]{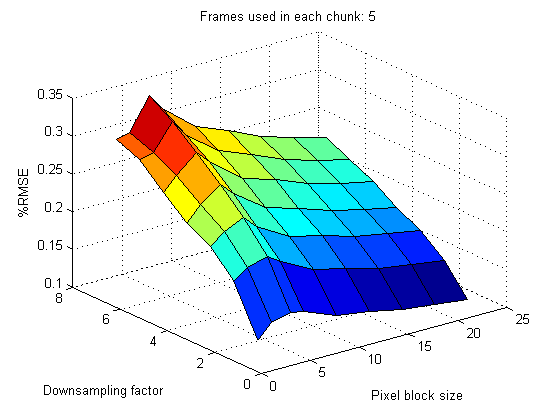}\includegraphics[width=1.2in]{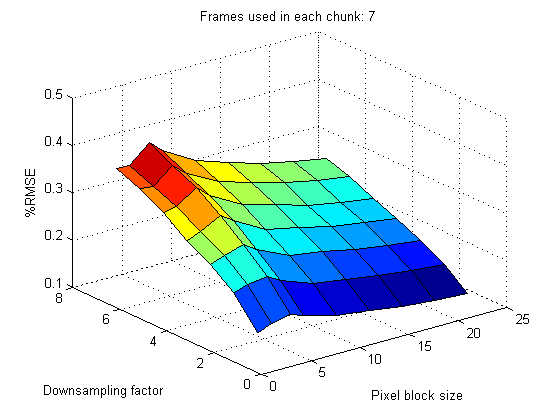}
\caption{Error (\%RMSE) of single Kronecker representation for 3, 5, and 7 frame blocks as a function of pixel block size and temporal downsampling factor.}
\label{Fig:E3}
\end{figure}

Figure \ref{Fig:E3} shows the Kronecker error for 3, 5, and 7 frame blocs as a function of downsampling factor and pixel block size.


\subsection{Human Activity Video}
For the computation of the sample covariance, we use a sliding window approach, where to obtain each new multiframe sample, the window is incremented by one frame. This in effect forces near block Toeplitz structure (for time stationarity) in the sample covariance.

We applied covariance estimation to the CMU human activity mocap videos. These videos are processed for the dataset to give the $(x,y,z)$ position of a set of fixed points as a function of time (downsampled to 40 frames/sec) on the human as the human performs an activity. We used videos of a person walking, and performing fencing moves. This type of data would also arise in situations where feature points in a video are being tracked.

The points travel through space, causing mean drift. To remove this, we preprocessed the data by computing the error of a $K$ frame ahead linear extrapolator (hereafter referred to as the zeroth order predictor) based on two frames close together in time. We then applied covariance estimation to the result. This setup allows for up to $K$ ahead causal prediction of the original variables.



\begin{figure}[htb]
\centering
\includegraphics[width=1.85in]{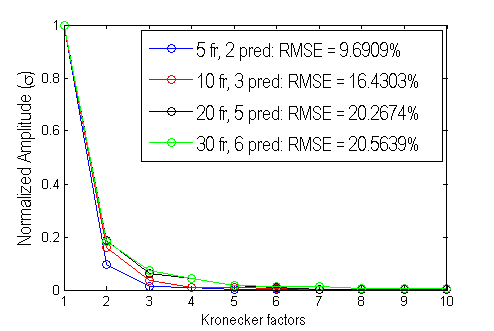}\includegraphics[width=1.85in]{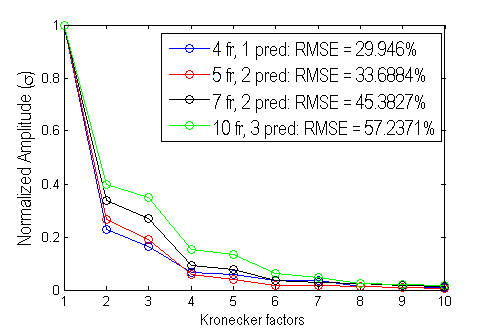}
\caption{Normalized RMS amplitudes of the first 10 terms of the LS sum of Kronecker products approximation to the covariance shown for a variety of covariance sizes. Also shown is the \%RMSE of using only the first Kronecker factor. 
Left: Fencing, 500 samples. Right: Walking, 100 samples. Note the concentration of energy in the first Kronecker factor.}
\label{Fig:CMU}
\end{figure}

In Figure \ref{Fig:CMU}, we show LS Kronecker covariance approximation results for CMU videos of fencing (44 $x,y,z$ points) and walking (downsampled to 14 points because of sample paucity). Approximations to a multiframe sample covariance learned using 500 and 100 samples respectively for the fencing and walking videos are considered. The RMS energy of the first 10 Kronecker product factors are shown for several different covariance sizes, as well as the \% RMSE of using only the first Kronecker product. The low number of samples for the walking video especially creates noise in the sample covariance, thus the RMSE values are somewhat inflated relative to the true covariance. 
%


\section{Results for Time Series Prediction}
\label{Sec:Results}
\subsection{Asymptotic CRB}
\begin{figure}[htb]
\centering
\includegraphics[width=3in]{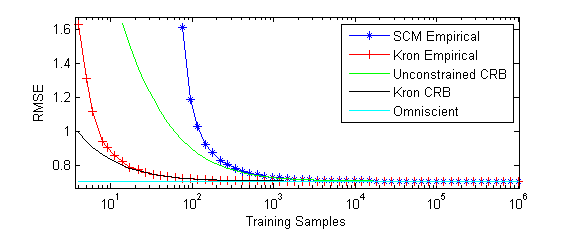}
\caption{Asymptotic prediction RMSE based on the predictor coefficient CRBs as a function of training sample size for Kronecker and standard covariance ML predictors, along with empirical performance curves. The generating covariance has a Kronecker form and was learned from the fencing video. The linear predictors are implemented by estimating the sample covariance matrix (SCM) and Kronecker product covariance, respectively, and using them to compute the ordinary least squares (OLS) prediction coefficients.} 
\label{Fig:CRB}
\end{figure}

Example CRB based asymptotic MLE prediction accuracy (found using Equation \eqref{eq:ErrCov}) results are shown in Figure \ref{Fig:CRB}, along with the Monte Carlo averages of prediction performance as a function of training sample size and the performance ($\mathrm{Cov}[y|x]$) achieved using perfect knowledge of the covariance (``omniscient"). The covariance matrix used was a Kronecker product ($\mathbf{T}\otimes \mathbf{S}$) LS approximation to a 7 frame covariance with 2 frame ahead prediction learned from the fencing video. The same covariance was used for both the sample covariance and Kronecker cases, with the only difference being that the Kronecker estimator has prior information that the covariance has Kronecker structure. As can be seen, our asymptotic CRB results match the asymptotic empirical performance well. 

\subsection{Forward Prediction}
\begin{figure}[htb]
\centering
\includegraphics[width=2.3in]{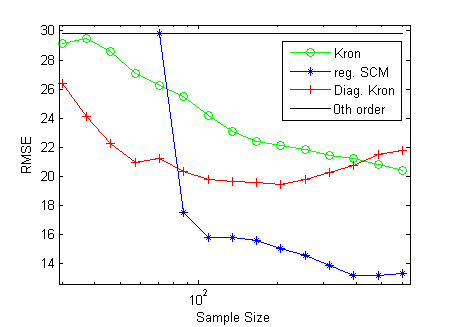}
\caption{CMU Fencing Video, prediction RMSE averaged over 100 frames as a function of learning sample size. Results shown for the zeroth order predictor, correction using regularized sample covariance, standard Kronecker LS approximation, and diagonally corrected Kronecker. Note the better performance of the Kronecker methods in the low sample regime. As sample size grows Kronecker bias begins to dominate and SCM outperforms the Kronecker models. Here the predictor was implemented with 10 frame covariance and 3 frame ahead prediction.} 
\label{Fig:ForPred}
\end{figure}

Figure \ref{Fig:ForPred} shows the RMSE results for forward prediction averaged over 100 consecutive frames in the CMU fencing video as a function of learning sample size. Prediction methods compared are the original predictor, the sample covariance (after L2 regularization \cite{tibshirani}), the standard LS Kronecker \eqref{KronApprox}, and the diagonally corrected Kronecker. Using the original predictor corresponds to using the mean (0) as the prediction, thus it can always be achieved using infinite regularization. In Figure \ref{Fig:ForPred}, the covariance is learned on the samples immediately prior to location at which prediction is occurring.

While most of the fencing video had strong enough Kronecker structure that additional Kronecker factors didn't improve prediction, some portions had sufficiently high Kronecker rank to warrant their use. By resampling from learned video covariances, we analyzed the RMSE as a function of the number of Kronecker factors used for both the standard \cite{TsiliArxiv} and diagonally corrected Kronecker methods. It was found that due to very poor conditioning, the standard Kronecker based predictions became unstable whereas the diagonally corrected estimate remained accurate when more than one Kronecker factor was used.

We used learned sample covariances from the fencing video and used it to generate new 10 frame sample covariances using the sliding window approach. The 5 frame ahead prediction RMSE was then computed over 200 Monte Carlo runs for the standard \cite{TsiliArxiv} and diagonally corrected Kronecker methods as well as the relearned sample covariance and true covariance as a function of the number of Kronecker factors used. Both Kronecker covariance estimates were forced to be positive semidefinite by projection. 

\begin{figure}[htb]
\centering
\includegraphics[width=1.6in]{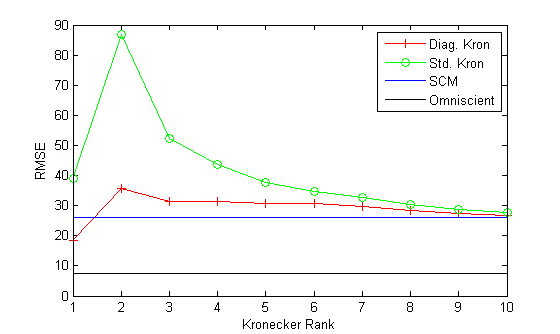}\includegraphics[width=1.6in]{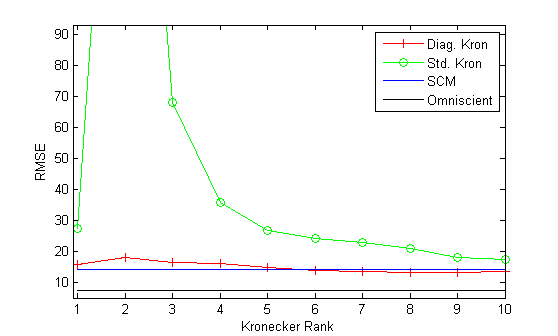}
\caption{Resampling from learned CMU Fencing Video covariance, prediction RMSE as a function of Kronecker approximation rank. 10 frame covariance, 5 frame ahead prediction. Results shown for the generating covariance (omniscient), sample covariance, standard Kronecker LS, and diagonally corrected Kronecker. Left: Results for 15 training samples. Right: Results for 200 samples.}
\label{Fig:RePred}
\end{figure}

Figure \ref{Fig:RePred} shows the results using 15 training samples (left) and 200 (right). Note the poor standard Kronecker results using more than one Kronecker term, demonstrating the benefit of diagonal correction in the multiple Kronecker case. 

\subsection{Forward Prediction with Partial Data}
An additional prediction task which may arise is forward prediction in the case that more recent history is available for some variables than for others. In the two part case we want
\begin{equation}
\hat{\mathbf{I}}_{1,t}\left|{\mathbf{\mu}},{\mathbf{\Sigma}},\{\mathbf{I}_{1,n}\}_{n=t-T+1}^{t-t_1}, \{\mathbf{I}_{2,n}\}_{n=t-T+1}^{t-t_2}\right.
\end{equation}
where $t_1 \neq t_2 \in [1,T]$, $\mathbf{I} = [\mathbf{I}_1 \: \mathbf{I}_2]$. 
The predictor (Equation \eqref{Eq:Predictor}) thus incorporates both ``forward" and ``sideways" prediction. For forward only prediction the structure of the single Kronecker model results in pixel predictions that are weighted averages of only the corresponding pixel values in the previous frames \cite{bonilla2008multi} with weights only a function of $\mathbf{T}$. 
In the partial prediction case this structure disappears resulting in the use of cross-pixel information. Since $\mathbf{S}$ is typically larger than $\mathbf{T}$, this results in an increase in the number of parameters. In addition, as discussed earlier, when uncorrelated noise is present the standard Kronecker has a tendency to overestimate inter-pixel correlations. Since the covariances we use have rather poor conditioning (large correlations) we expect the predictions using the standard Kronecker estimate to degrade significantly even for large numbers of samples and that the diagonally corrected method will result in better performance.

\begin{figure}[htb]
\centering
\includegraphics[width=2.2in]{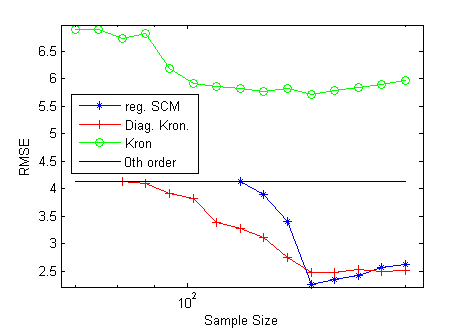}
\caption{Walking Video. Partial data prediction RMSE as a function of training samples using the standard Kronecker, diagonally corrected Kronecker, and regularized sample covariance predictors, and the zeroth order predictor. Unregularized SCM is not shown due to excessive magnitude.}
\label{Fig:PartPred}
\end{figure}

For this experiment, we used the walking video from the CMU dataset. Figure \ref{Fig:PartPred} shows the RMSE averaged over 100 frames of predicting 2/3 of the points (1/3 are observed at all times) 5 frames ahead using a 20-frame covariance. Note the failure of the standard Kronecker as anticipated, while the diagonally corrected Kronecker has lower error in the low sample regime than the regularized sample covariance.

\section{Conclusion}
\label{Conclusion}
We considered the sum of Kronecker products representation for covariance matrices developed in \cite{TsiliArxiv}, and examined its applicability to spatio-temporal covariance estimation, especially in video. It was found that a small sum (usually two) of Kronecker factors is a good approximation to the covariance of the CMU videos, and due to the reduction in the number of parameters gives improved low-sample estimation as compared to the standard sample covariance matrix. 

We also proposed a diagonally loaded sum-of-Kronecker products representation, which resulted in improved prediction performance. As an example application, we used it for prediction of human motion patterns in the CMU videos. In addition to the significant potential computational improvement of using a single Kronecker factor, it was found that the representation allowed accurate prediction using significantly fewer training samples than needed using the sample covariance matrix.

To analyze this gain, we derived the CRB for the predictor coefficients and the optimal asymptotic predictor performance assuming an underlying Kronecker covariance as well as for the unstructured covariance case. 

In certain cases the use of multiple Kronecker factors using the diagonally corrected method gave improved performance as the number of samples increased sufficiently, whereas the standard method gave worse performance. This allowed the small sum-of-Kroneckers representation to be competitive even for large numbers of samples.

%
%
%

\section{Acknowledgements}
This research was partially supported by ARO under grant W911NF-11-1-0391 and by AFRL under grant FA8650-07-D-1220-0006. The CMU data was obtained from mocap.cs.cmu.edu. The dataset was created with funding from NSF EIA-0196217.

\bibliographystyle{IEEETran}
\bibliography{CAMSAP_bib}

\end{document}